\documentclass[%
 reprint,
preprintnumbers,
 amsmath,amssymb,
 aps,
prb,
]{revtex4-1}

\usepackage{graphicx}
\usepackage{dcolumn}
\usepackage{bm}
\usepackage{listings}
\usepackage{subcaption}
\usepackage{multirow}

\begin{document}
\preprint{\large SAND2020-0316 O}

\title{Many-body electronic structure of LaScO$_3$
by real space quantum Monte Carlo}

\author{Cody A. Melton}
\email{cmelton@sandia.gov}
 \affiliation{High Energy Density Physics Theory, Sandia National Laboratories}
\author{Lubos Mitas}%
 \email{lmitas@ncsu.edu}
\affiliation{
Department of Physics, North Carolina State University
}%





\date{\today}

\begin{abstract}
We present real space quantum Monte Carlo (QMC) calculations of the scandate
LaScO$_3$ that proved to be challenging for traditional electronic structure approaches due to
strong correlation effects resulting in inaccurate band gaps from DFT and $GW$ methods when compared with existing experimental data.
Besides calculating an accurate QMC band gap corrected for supercell size biases and in agreement with numerous experiments, we also predict the cohesive energy of the crystal using the standard fixed-node QMC
without any empirical or non-variational parameters. We show that promotion (optical) gap and fundamental gap agree with each other illustrating a clear absence of significant excitonic effects in the ideal crystal. We obtained these results in perfect
consistency in two independent tracks that employ different basis sets
(plane wave vs. localized gaussians), different codes for generating orbitals
(\textsc{Quantum Espresso} vs. \textsc{Crystal}), different QMC codes
(\textsc{Qmcpack} vs. \textsc{Qwalk}) and different high-accuracy pseudopotentials
(ccECPs vs. Troullier-Martins) presenting the maturity and consistency of QMC methodology and tools for studies of strongly correlated problems. 
\begin{description}
\item[PACS numbers]
May be entered using the \verb+\pacs{#1}+ command.
\end{description}
\end{abstract}

\pacs{Valid PACS appear here}
\maketitle
\section{Introduction}
Transition metal oxides that crystalize in perovskite structures are intensely studied materials due to their complex many-body effects as well as for their potential for broad applications. 
The challenges in understanding their electronic properties are well-known and include electron-electron correlation, competition of localized $d$ vs. more extended $s$ and $p$ states, crystal field effects, and spin-orbit contributions for heavier elements. 
A number of studies over the decades have been devoted to understand these effects 
within a range of approaches such as density functional theory (DFT), post-DFT, and dynamical mean-field theory (DMFT). 
Very recently, a group of La-based perovskites with formula LaMO$_3$ where M represents a $3d$ transition metal \cite{PhysRevB.51.16575,PhysRevB.86.235117,PhysRevMaterials.2.024601,doi:10.1063/1.4994083}
have proved to be rather challenging even for the most advanced methods.  
 Most of these La-based crystals show local magnetic moments and magnetic orderings as is often the case for 
 transition element perovskites with partially filled shells, especially for the middle/late $3d$ series.
Interestingly, the scandate LaScO$_3$ is rather atypical since the scandium $d$-shell becomes nominally empty and the ground state therefore appears to be nonmagnetic. 
The La atom contributes charge towards closing the empty shells of oxygens and stabilizes the structure by filling the empty space between the ScO$_6$ octahedra in the typical perovskite arrangement and results in a large gap, as determined by various optical experiments \cite{doi:10.1063/1.1829781,PhysRevB.48.17006,Heeg2006}.
Seemingly, its electronic structure should be not much different from other nonmagnetic closed-shell perovskites that have been described with a mixed degree of accuracy by a variety of post-DFT approaches such as $GW$, DFT+$U$, and more elaborate DMFT methods. 
It is therefore somewhat unexpected that this system proved to be difficult to reconcile with available experimental data, since the gap estimations are smaller than experiments by 1 eV or more, even in post-DFT methods.
As explained by Millis et al \cite{PhysRevB.90.125114}, the arbitrary boost to Hubbard $U$, which is often used as a fix for 
ordinary DFT methods to open the gap in transition
metal oxides, is not helpful since the $d$-levels are nominally unoccupied. 
Indeed, {\it ad hoc}, educated double-counting schemes adapted for DFT+DMFT had to be employed in order to build the theoretical understanding of experimental gaps so that a proper  balance of the charge-transfer vs. Mott-like behavior is reached in related systems such as LaTiO$_3$ \cite{PhysRevB.90.125114}.
Alternative methods such as $GW$ show somewhat mixed success for this class of systems. 
In a systematic and very precise study with several variants of the $GW$ approach \cite{PhysRevMaterials.2.024601}, heavier transition metal oxide gaps agree quite well with the experiment,
whereas in LaTiO$_3$, the gap is overestimated by $\sim 0.9$~eV. 
On the other hand for  LaScO$_3$, the study shows an underestimated gap by $\sim 1-1.5$~eV, depending on the details of 
the $GW$ method. 
For LaScO$_3$ in particular, the resulting discrepancy has led to the claim that the $GW$ result is more reliable than the experiment in this case, despite numerous optical studies indicating the gap is at least as large 
as 5.7~eV, and quite possibly  larger. 
Considering these contradictions, here we present an independent many-body study of LaScO$_3$ system using real space quantum Monte Carlo which has paved the way for employing correlated wave function electronic structure methods to real materials with strong correlations over the last two decades. 
We find agreement with the experimental band gap without any empirical inputs or {\it ad hoc} parameter tuning, merely using the fixed-node diffusion Monte Carlo (FNDMC) approach in its variational formulation, and corroborate our result with two independent QMC codes.
We probe the band gap using the canonical definition as well as optical promotion excitation with the same results. We also propose a method for analysis and correction of finite size effects for the band gap estimation with increasing supercell sizes. 
In addition, we {\em predict} the value of the crystal cohesion energy, for which to the best of our knowledge there is no experimental determination. Therefore, this represents a genuine prediction from our many-body methods. 
In addition, we find that despite being a strongly correlated system, LaScO$_3$ is very well described by a single-reference wave function nodal surface with corresponding correlations recovered by the ordinary DMC method
leading to accurate energy differences.

\section{Methods}
\begin{figure}[!t]
  \centering
  \includegraphics[width=0.4\textwidth]{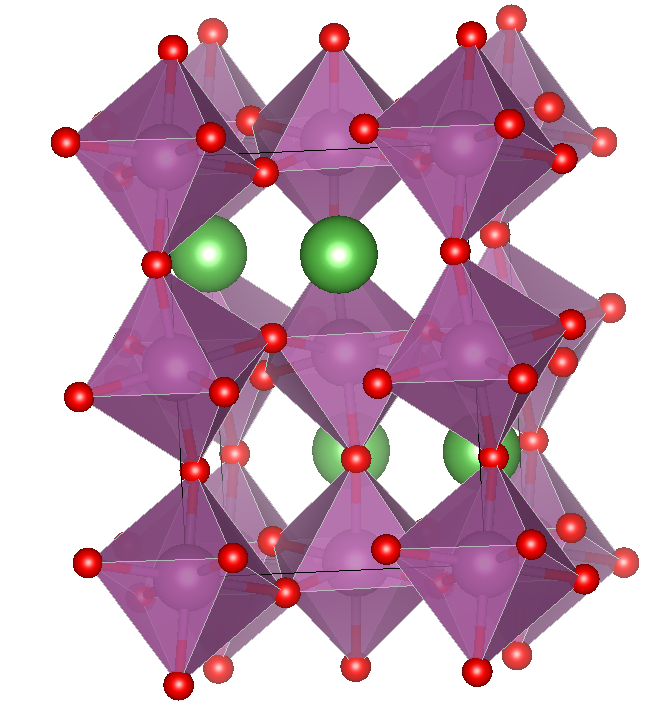}
  \caption{LaScO$_3$ in the experimental perovskite structure. ScO$_6$ octahedra
    are shown in purple, with the O atoms shown in red. The La$^{+3}$ cations
    are shown in green. This figure was generated using the VESTA software \cite{vesta}.
  }
  \label{fig:structure}
\end{figure}

ABO$_3$ perovskites, with A, B as cations and O being the anion, can crystallize
in a variety of crystal symmetries. LaScO$_3$ in particular crystallizes with the $Pnma (62)$ crystal
symmetry.  Oxygen anions surround the Sc
cations to form ScO$_6$ octahedra, while the La cations fill the space between
these octahedra and close the O $p$ shell. Since Sc$^{+3}$ and La$^{+3}$ are closed-shell, LaScO$_3$ is
nonmagnetic as opposed to other LaBO$_3$ perovskites, where the partially filled
$d$ shells facilitate G-type anti-ferromagnetism. 
The experimental structural information is given in Table \ref{tab:structure} and illustrated in Fig. \ref{fig:structure}. 
We note that
while the previous DFT and post-DFT calculations that we will be comparing to
have used a slightly different experimental structure
\cite{Geller:a01971}, we use a more recent experimental determination of the
structure \cite{CLARK1978129}. The difference between these experimental
structures results in only a 0.05\% difference in the volumes and
less than 1\% error for the various Sc-O octahedral bonds. Hybrid DFT studies with HSE06
find that the DFT relaxed structures all have similar percent
errors to the older experimental structure, and that the band gap is sensitive to at most 0.1~eV between the
experimental and relaxed geometries \cite{PhysRevB.86.235117}. Therefore, while
all calculations considered in this work use the more recent experimental
geometry, any comparison to DFT and post-DFT methods that use the older
experimental geometries would change the differences only marginally, within approximately 0.1~eV, if
our adopted geometry had been used. 
\begin{table}[!t]
  \centering
  \caption{Experimental structure for LaScO$_3$ in the $Pnma (62)$ crystal
    structure \cite{CLARK1978129}.}
  \label{tab:structure}
  \begin{tabular}{cc|ccc}
    \hline\hline
    \multicolumn{2}{c|}{\multirow{2}{*}{Lattice Parameters}} & $a$ (\r{A}) & $b$ (\r{A}) & $c$ (\r{A})\\
    \cline{3-5}
                                          & & 5.7911 & 8.0923 & 5.6748\\
    \hline
    Atom & Site & $x$ & $y$ & $z$ \\
    \hline
    O$_1$ &8$d$ & 0.1988 & 0.0528 & 0.3044 \\
    O$_2$ &4$c$ & 0.43745 & 1/4 & 0.01537 \\
    La &4$c$ & 0.5323 & 1/4 & 0.5996 \\
    Sc &4$a$ & 0 & 0 & 0 \\
    \hline
    \hline
  \end{tabular}
\end{table}

As opposed to effective Hamiltonian methods like DFT, DFT+$U$, DFT+$GW$, DFT+DMFT, etc., diffusion Monte Carlo (DMC) works directly with the many-body wave function. 
Using the imaginary time Schr\"{o}dinger equation, we apply an imaginary time evolution operator to a trial wave function $e^{-\tau H} \Psi_T$ which evolves to the ground state in the long time evolution limit. 
Although this is formally exact, in order to deal with the fermion sign problem,
we employ the well-known fixed-node approximation (FNDMC) \cite{Foulkes2001}.
In this 
approximation, the obtained solution is enforced to have the same zero/nodal surface as the trial wave function. 
This amounts to adding an additional term to the Hamiltonian, $H \rightarrow H + V_\infty \delta(\mathbf{R}-\mathbf{R}_{\partial\Omega})$, where $\mathbf{R}_{\partial\Omega} = \left\{ \mathbf{R}; \Psi_T(\mathbf{R}) = 0 \right\}$ while
$\mathbf{R}$ denotes coordinates of all involved electrons. The strength of the interaction is such that $V_{\infty} \to\infty$, so that it enforces a zero boundary condition at the trial nodal surface
\cite{CodyPRE}. 
Since a non-exact nodal surface can only raise the energy relative to the exact energy, the expectation value of such Hamiltonian is therefore variational. 

The accuracy of the method clearly relies on the accuracy of the trial wave function. 
Throughout this work, we utilize a Slater-Jastrow wave function
\begin{equation}
    \Psi_T(\mathbf{R}) = e^{J(\mathbf{R})}D_\uparrow(\psi_k^\uparrow(\mathbf{r}_i))D_\downarrow(\psi_l^\downarrow(\mathbf{r}_j)) \,.
\end{equation}
Here, $D_\sigma$ is a Slater determinant for spin $\sigma$ and $\psi_i^\sigma(\mathbf{r}_j)$ form the single-particle orbital matrix. 
The Jastrow factor $J(\mathbf{R})$ builds explicit correlation into the trial wave function by considering electron-electron, electron-ion, and electron-electron-ion terms; clearly, the Jastrow cannot modify the nodal surface. 
In order to carry out a restricted optimization of the nodal surface, we use various sets of single-particle orbitals obtained from Hartree-Fock (HF), LDA, and PBE0$_w$ where $w$ determines the amount of exact exchange in PBE0, defined as
\begin{equation}
    {\rm PBE0}_w = w E^{\rm HF}_{\rm x} + (1-w) E^{\rm PBE}_{\rm x} + E^{\rm PBE}_{\rm c} \,.
\end{equation}
For both the ground and excited state calculations considered, we optimize the
Jastrow coefficients within variational Monte Carlo (VMC) using the linear
method \cite{PhysRevLett.98.110201}, and take the lowest energy from FNDMC as the
optimal nodal surface.

In order to estimate the gap within DMC, tradionally two approaches have been utilized, namely the {\it quasiparticle} and {\it optical} formulations \cite{PhysRevB.98.075122,yang2019electronic,frank_many-body_2019}.
In the quasiparticle approach, the gap is estimated by considering electron addition/removal from the wave function.
By considering the addition/removal of an electron from the valence band maximum (VBM) and the conduction band minimum (CBM), we directly target the fundamental gap of the material. It is given by the following difference of total energies for system with $(N\pm 1)$ and $N$ electrons 
\begin{eqnarray}
  \Delta_{\rm qp} &=& E_{N+1} + E_{N-1} - 2E_N \nonumber \\
    \Psi_{N-1} &=& c_{\rm VBM}\Psi_N\nonumber  \\
    \Psi_{N+1} &=& c^\dagger_{\rm CBM} 
                               \Psi_{N} \,,
                               \label{eqn:fundamental}
\end{eqnarray} 
with $\Psi_{N(\pm 1)}$ being the eigenstate for the $N(\pm 1)$ electron
systems. 
Since each addition/removal is intended to be the ground state of the respective
$N\pm 1$ systems, the FNDMC variational theorem holds and we take the lowest
energy nodal surface.  
An alternative method is to use the optical excitation approach, which is argued to target
the optical gap instead of the fundamental gap, potentially including excitonic
information due to the electron-hole interaction. 
This is achieved by promoting a particle to an excited state, typically from the valence band maximum  to the conduction band minimum 
\begin{eqnarray}
    \Delta_{\rm op} &=& E^{\rm ex}_N - E_N  \nonumber \\
    \Psi_N^{\rm ex} &=&  c^\dagger_{\rm CBM}
                                  c_{\rm VBM} \Psi_N 
                                  \,,
\end{eqnarray}
where $\Psi^{\rm ex}$ is the wave function for the excited state system. Although in general, the fixed-node approximation is only variational for the
ground state, often the change of symmetry from VBM to CBM state guarantees the orthogonality of the excitation and the variational bound. Even when
both VBM and CBM are of the same symmetry, the nodal constraint is sufficiently restrictive that single-particle
excitations results in variational behavior in practice  \cite{PhysRevB.98.075122}, (see Foulkes {\it et al.}  \cite{foulkes_exc_symm} for possible exceptions).
In some cases, the optical gap may not be adequately described by the single-particle excitation used in the optical approach and a more elaborated methodology to target excited states has been developed \cite{zhao19} for QMC methods; however, here we restrict ourselves to the common single particle-excitation due to the absence of any pronounced excitonic effects.  
In cases where the dielectric constant is large, excitonic effects are largely screened out, and thus the optical and quasiparticle approach are nominally equivalent. This should be indeed the case
for LaScO$_3$ with a dielectric constant of $\epsilon \simeq 24$ \cite{HEEG2005150}, therefore we expect both approaches to produce equivalent results within the energy resolution of our QMC calculations. 

In order to test the validity of our calculations, we utilize two independent QMC codes, namely \textsc{Qwalk} \cite{WAGNER20093390} and \textsc{Qmcpack} \cite{Kim_2018}.
For the \textsc{Qwalk} calculations, we generate our trial wave functions using orbitals
with localized gaussian basis sets with \textsc{Crystal} \cite{CRYSTAL,CRYSTAL09}.
Gaussian orbital codes like \textsc{Crystal} are well-suited for semi-local effective core potentials, which allows us to use our recently generated correlation consistent effective core potentials (ccECPs) for Sc \cite{doi:10.1063/1.5040472} and O \cite{doi:10.1063/1.4995643}, which are designed for explicitly correlated methods like FNDMC. 
Since at present a ccECP for La has not been constructed, we instead use an existing ECP
from the Stuttgart group \cite{Dolg1989,Dolg1993}, which has been appropriately
modified to remove the Coulomb singularity in the potential, without changing
the properties of the potential or any energy differences. 
We generate basis sets of the TZDP quality for O$^{-1}$, La$^{+2}$ and Sc$^{+2}$,
which are close to the realized oxidation states in LaScO$_3$. 
In order to utilize our ccECPs within \textsc{Qmcpack}, periodic boundary conditions are mostly supported by orbital generation from \textsc{Quantum Espresso} (QE) \cite{QE-2009,QE-2017}.
This requires a transformation of our semi-local ccECPs to the non-local Kleinman-Bylander (KB) form \cite{PhysRevLett.48.1425} for orbital generation, although the original semi-local pseudopotential is used in the actual QMC calculations. 
While this is an exact transformation for the reference state, in general the potential is different than the semi-local form and can introduce errors such as ghost states and/or
compromised transferability in subsequent QMC calculations \cite{PhysRevB.94.165170}.
We found that while our ccECPs for Sc and O did not result in ghost states, the Stuttgart group pseudopotential did, and therefore we utilized another pseudopotential for La designed in DFT
that has been specifically adapted to to be used in Kleinman-Bylander (KB) form with plane wave basis set \cite{PhysRevB.41.1227}. We denote this 
La pseudopotential as dft-kb. 
By comparing \textsc{Qmcpack} and \textsc{Qwalk} calculations, we cross-check that quality of the KB transformation of our ccECP potentials for Sc and O. 
In addition, we also test another set of pseudopotentials for Sc and O designed for QMC calculations
\cite{PhysRevB.93.075143} which we denote as dft-opt and compare the results with our ccECPs using \textsc{Qwalk}.


\section{Results}
\subsection{Optimal nodal surfaces}
First, we focus on finding optimal nodal surfaces using orbital sets generated 
by hybrid DFT function PBE0$_w$ where 
$w$ denotes the weight of the exact exchange (the original PBE0 functional has $w=0.25$).
\begin{figure*}[!ht]
  \centering
    \begin{subfigure}
        {0.3\textwidth}
        \centering
        \includegraphics[width=\textwidth]{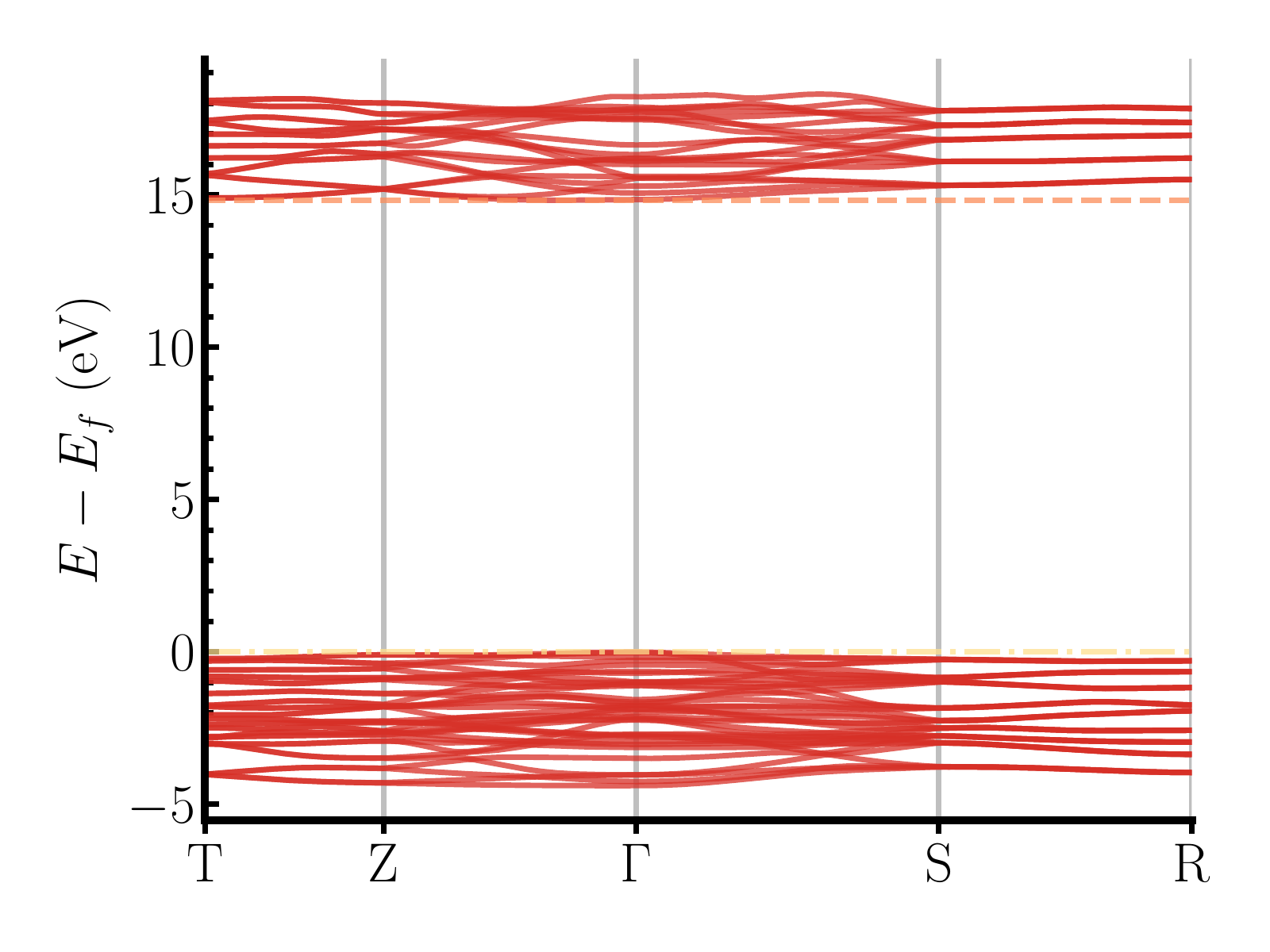}
       \caption{HF}
        \label{fig:hf}
    \end{subfigure}
    \begin{subfigure}
        {0.3\textwidth}
        \centering
        \includegraphics[width=\textwidth]{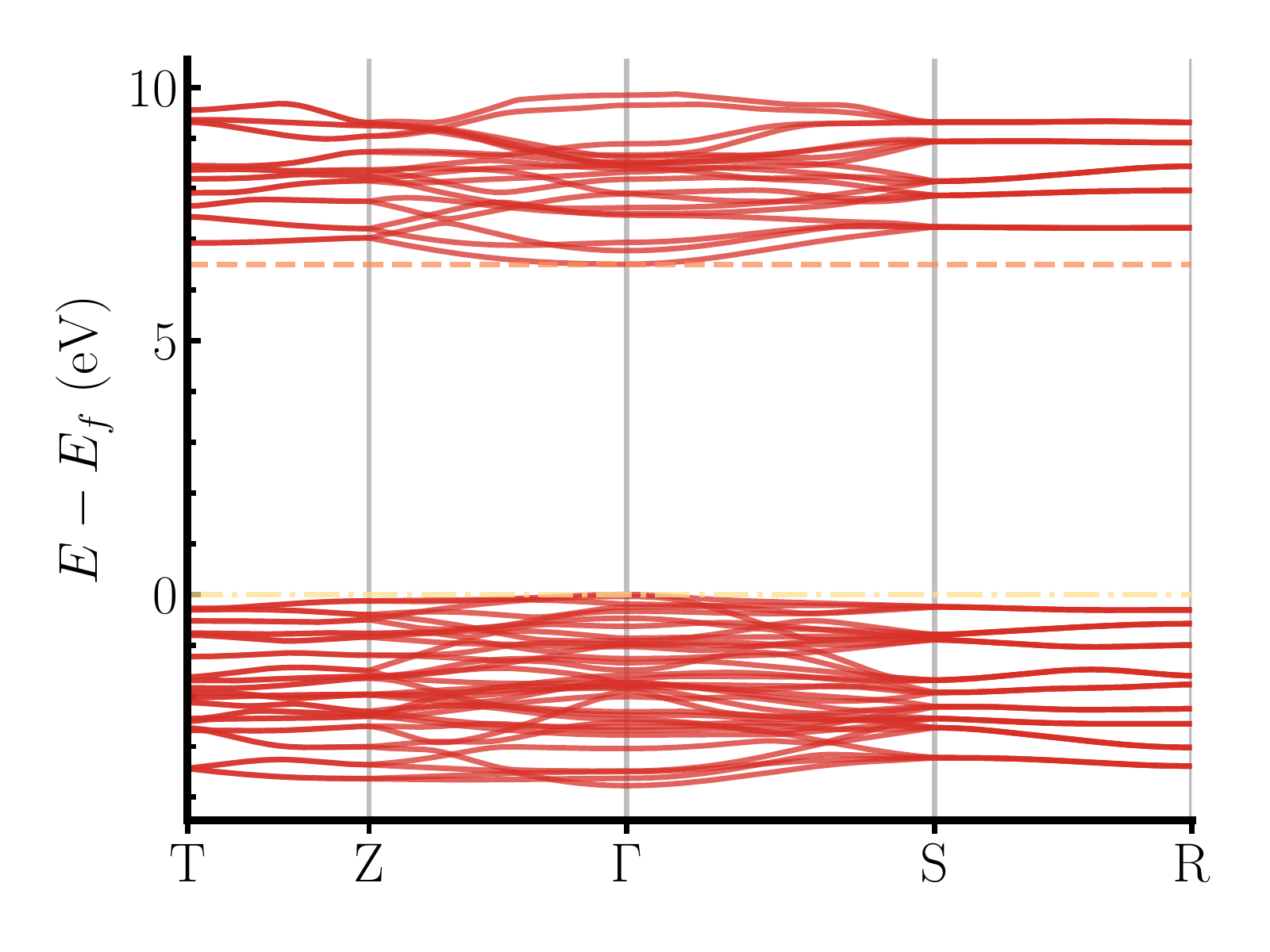}
        \caption{PBE0}
        \label{fig:pbe0}
    \end{subfigure}
    \begin{subfigure}
        {0.3\textwidth}
        \centering
        \includegraphics[width=\textwidth]{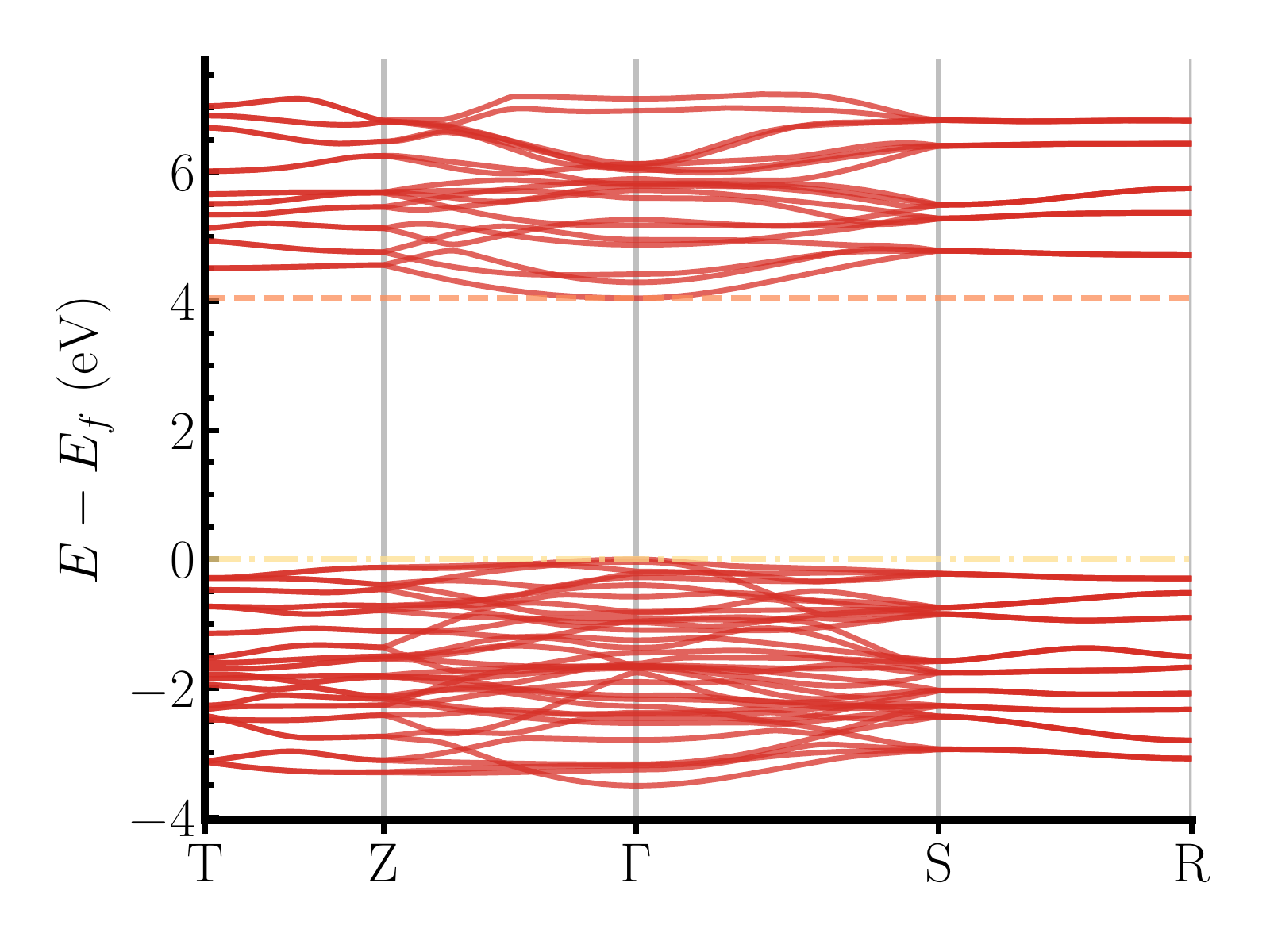}
        \caption{PBE}
        \label{fig:pbe}
    \end{subfigure}
    \caption{LaScO$_3$ self-consistent band structures in the experimental geometry with varying degree of exact exchange. (\subref{fig:hf})  Hartree-Fock (HF) band structure gap of 14.82~eV is indirect with the valence band minimum (VBM) at $\Gamma$. However, the conduction band is very flat with a marginal difference between the precise conduction band minumum (CBM) and the $\Gamma$ point. (\subref{fig:pbe0}) The PBE0  band structure shows a direct gap of 6.51~eV at $\Gamma$. Similarly, (\subref{fig:pbe}) the PBE band structure shows a direct gap of 4.04~eV at $\Gamma$.}
    \label{fig:bandstructure}
\end{figure*}
In order to see the impact of nodal changes  both on the ground and excited states, we 
probe the optical excitation where we simply promote an electron in the Slater determinant from VBM to CBM. For this purpose, we plot  
 the one-particle band structures from Hartree-Fock (HF), PBE0 and PBE calculations in Fig.
\ref{fig:bandstructure}. The plot shows  that the band extremes are both at the $\Gamma$ point in the range of $w$ where we expect the optimal mixing of the exact exchange $w$ that typically falls between
10 and 40\% \cite{WAGNER2003412,kolorencprb}.

\begin{figure*}[!t]
  \centering
    \begin{subfigure}
        {0.4\textwidth}
        \centering
        \includegraphics[width=\textwidth]{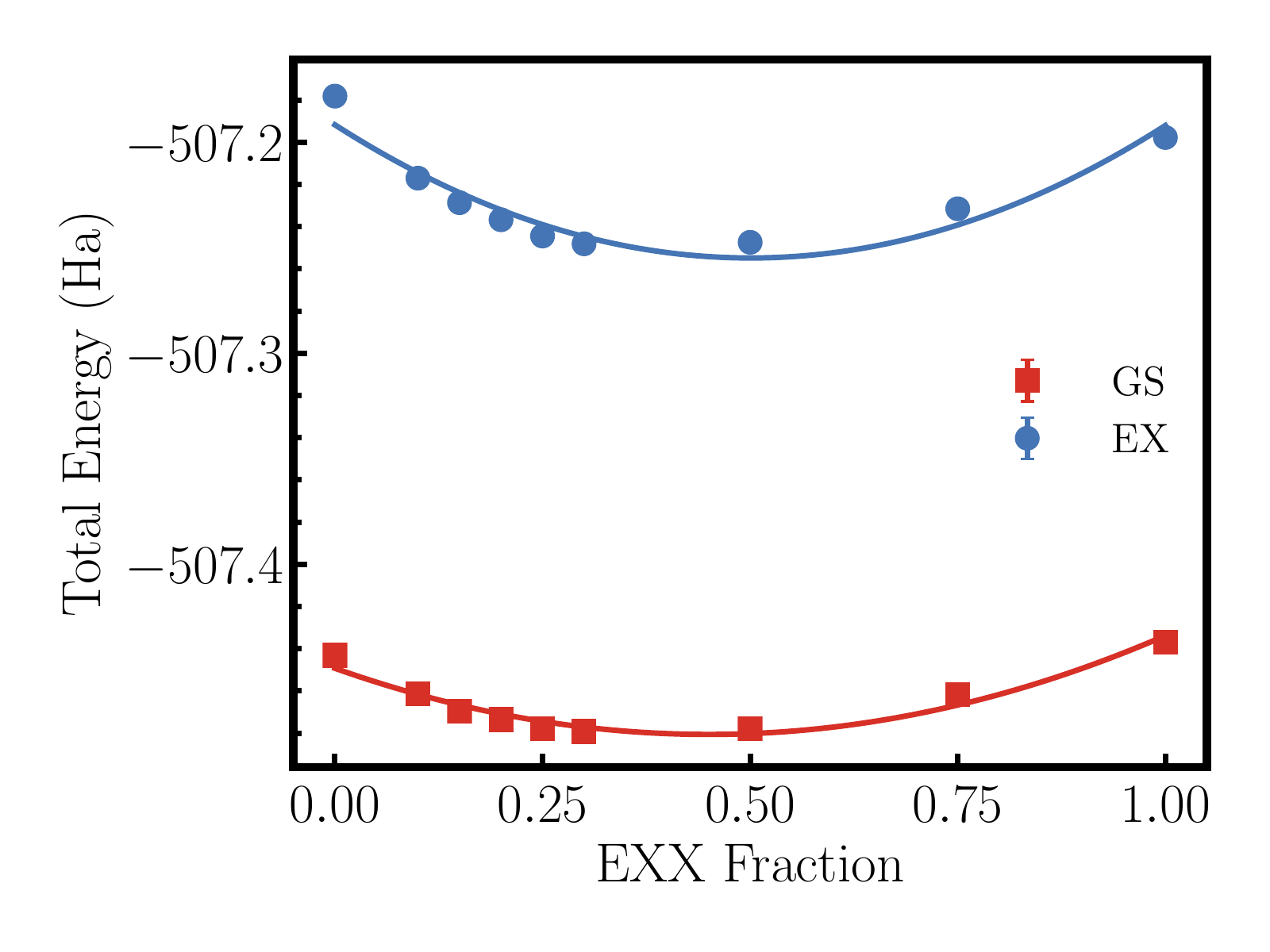}
        \caption{FNDMC with ccECPs and dft-kb for La}
        \label{fig:qmcpack_ccECP}
    \end{subfigure}
    \begin{subfigure}
        {0.4\textwidth}
        \centering
        \includegraphics[width=\textwidth]{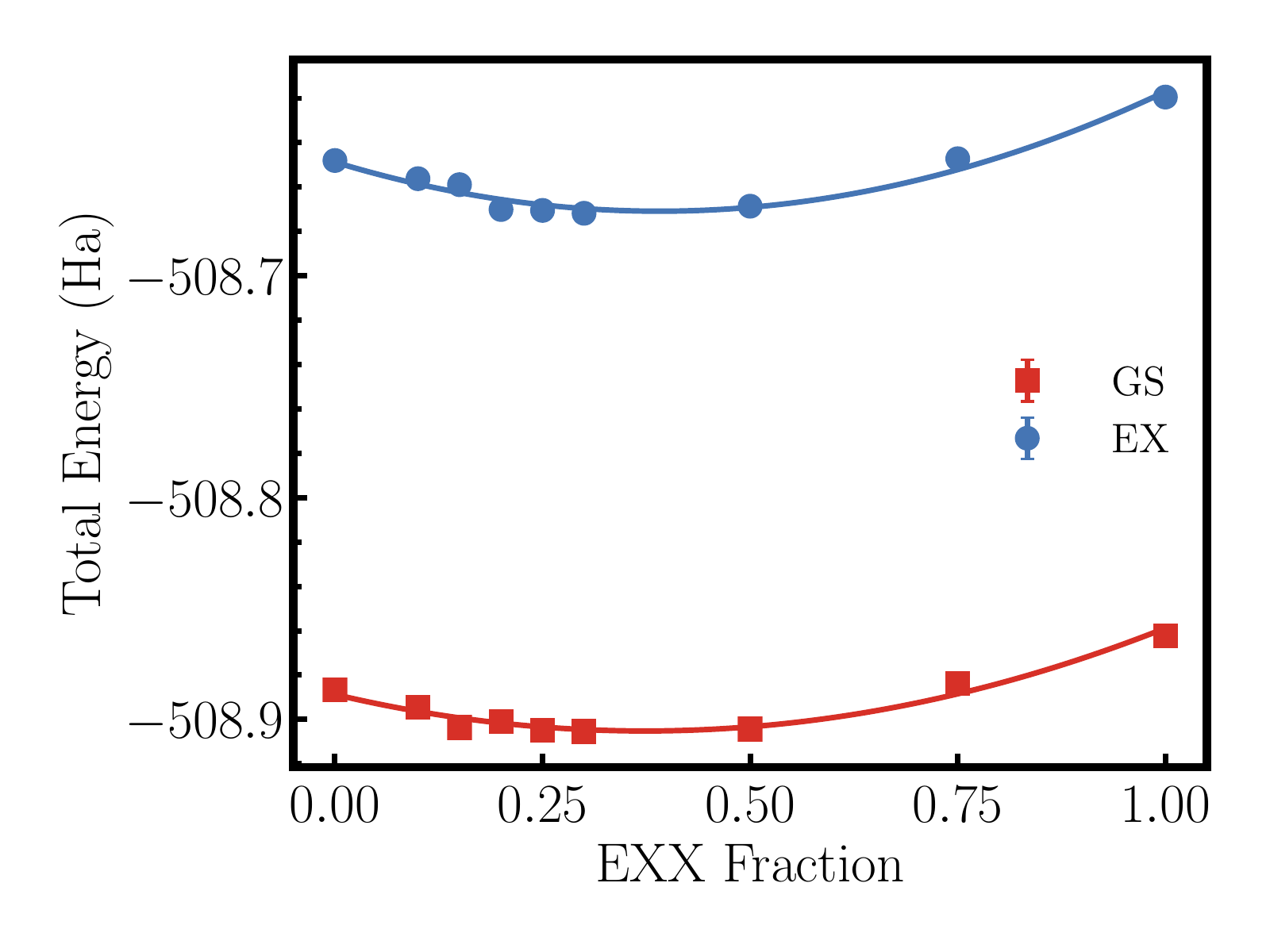}
        \caption{FNDMC with dft-opt pseudopotentials}
        \label{fig:qmcpack_jaron}
    \end{subfigure}
    \caption{FNDMC total energies of  LaScO$_3$ ground (squares) and excited (circles) states at the $\Gamma$ point plotted as functions of the exact exchange weight $w$.
    The corresponding orbital sets are generated by PBE0$_w$ and FNDMC calculations are carried out using \textsc{Qmcpack}. Part (\ref{fig:qmcpack_ccECP}) uses ccECPs for Sc and O, and  dft-kb pseudopotential for La
    \cite{PhysRevB.41.1227}.
     Part (\ref{fig:qmcpack_jaron}) is  calculated with dft-opt pseudopotentials \cite{PhysRevB.93.075143}.}
    \label{fig:qmcpack_gaps}
\end{figure*}

In Fig. \ref{fig:qmcpack_gaps}, we show the calculated ground and excited states
from DMC using the \textsc{QMCPACK} code.
These calculations use a 20-atom simulation cell, with 160 electrons, and we perform
the ground state and excited state calculations with a $\Gamma$ point twist
vector.

By mixing the hybrid exchange in the density functional used to generate the Slater part of the determinant, we are able to optimize the nodal surface for both the ground state and excited state. 
Fig. \ref{fig:qmcpack_ccECP} shows the calculations using the ccECPs for Sc and O and dft-kb
pseudopotential for La \cite{rappe}, whereas the Fig. \ref{fig:qmcpack_jaron} shows the same for the dft-opt pseudopotentials.
Irrespective of the pseudopotentials used, the total energies for the ground and excited state have similar behavior. The minima occur around $w=0.3$ mixing although 
for the ground states PBE0 with $w=0.2,0.25$ are essentially identical by being basically within the error bar.

\subsection{Cohesive Energy}
In order to further test the quality of the softer dft-opt \cite{PhysRevB.93.075143} pseudopotentials generated
for use in QE and \textsc{Qmcpack}, we also directly compare the cohesive energies using two independent strands of calculations with different ECPs, basis sets and QMC codes.
In particular, we utilize our ccECPs for Sc and O, Stuttgart ECP for La, in the \textsc{Crystal} code for generating the orbitals expanded in gaussian basis sets and running the FNDMC with \textsc{Qwalk}. The other independent track of calculations uses dft-opt pseudopotentials, Quantum Espresso with plane waves and FNDMC calculations carried out with \textsc{Qmcpack}.
The atomic energies are calculated with single-reference fixed-node trial wave
functions, and can be found in Table \ref{tab:atoms}. These will be used as a
reference when calculating the cohesive energy of LaScO$_3$. 

We perform twist averaging to deal with the single-particle finite size effects,
both with a variety of twist grids shown in Table \ref{tab:coh}.
Each twist averaged calculation is carried out using the nodal surface generated
by PBE0$_{25}$ nodes.
We point out that if we consider the $\Gamma$ point only, we find excellent
agreement between the cohesive energies of the two
sets of results. 
This is quite encouraging, given the difference in orbital generation, QMC codes, effective Hamiltonians
and different basis sets. 
In addition, as we converge the twist grids to the 4x4x4, we obtain again excellent agreement.
The disagreement at intermediate twists in due to symmetry considerations when
generating the one-particle orbitals.
Given the agreement between the two QMC codes and various pseudopotentials, we focus on the \textsc{Qwalk} results for extrapolation to the thermodynamic limit.
The extrapolation is used to filter out the two-body finite size effects and to this end, we perform a supercell calculation and
extrapolate to the thermodynamic limit alongside the  twist-averaging.  The supercell is chosen such that it
doubles the number of atoms and it is as close to 
a cube as possible. The cohesive
energy per chemical formula is typically linearly extrapolated to the thermodynamic limit with scaling variable  $1/N$ where $N$ is the number of formula units.
The various
cell energies are collected in Table \ref{tab:cohesion_super}.
Note that already at 40 atom supercell the energy components are close to 
the twist averages.
When extrapolated to the thermodynamic limit, they yield cohesive energy of 1.329~Ha per formula unit. 
To the best of our knowledge, there is no experimental
cohesion available, and thus this is a genuine prediction.

\begin{table}[!t]
  \centering
  \caption{Atomic energies from fixed-node diffusion Monte Carlo and single reference
    single-reference trial wave functions. For the \textsc{Qwalk} calculations, we
    utilize our ccECPs and the dft-kb pseudopotential for La, while for the \textsc{Qmcpack} calculations here we use
    dft-opt pseudopotentials.}
  \label{tab:atoms}
  \begin{tabular}{c|cc}
    \hline\hline
    Atom          & \textsc{Qwalk}/ccECP &  \textsc{Qmcpack}/dft-opt \\
    \hline
        La        &  -31.40293(22) &  -31.536684(4)  \\
        O         &  -15.86548(39) &  -15.90088(27)  \\
    Sc            &  -46.52048(35) &  -46.63115(48)  \\
    \hline
    \hline
  \end{tabular}
\end{table}

\begin{table}[!t]
    \centering
    \caption{Comparison of the twist-averaged energies for the 20-atom
      simulation cell between \textsc{Qwalk} and \textsc{Qmcpack}. The \textsc{Qwalk} calculations utilize
      the ccECPs (dft-kb for La), whereas the \textsc{Qmcpack} calculations are using dft-opt pseudopotentials. The corresponding atomic energies are in Table \ref{tab:atoms}. Total energies are given per formula unit.}
    \label{tab:coh}
    \begin{tabular}{c | c c | c c }
        \hline\hline
             & \multicolumn{2}{c}{\textsc{Qwalk}/ccECP} & \multicolumn{2}{c}{\textsc{QMCPACK}/opt}\\
        \hline
        Twists    & E (Ha) & Coh. (Ha) & E (Ha) & Coh. (Ha) \\
        \hline
        $\Gamma$  & -126.87466(6) & 1.3548(08) & -127.2262(04) & 1.3558(08)  \\
        2x2x2     & -126.8903(08) & 1.3704(11) & -127.2568(08) & 1.3864(10) \\
        3x3x3     & -126.8956(10) & 1.3757(10) & -127.2546(04) & 1.3841(07)  \\
        4x4x4     & -126.9058(05) & 1.3860(09) & -127.2556(02) & 1.3851(07)  \\
        \hline
        \hline
    \end{tabular}
\end{table}

\begin{table}[!t]
  \centering
  \caption{Total, kinetic, and potential energies for the various \textsc{Qwalk}
    calculations. This includes two supercell sizes and several
    twist grids. Energies are given per formula unit.}
  \label{tab:cohesion_super}
  \begin{tabular}{cc|ccc}
    \hline\hline
    $N_{\rm at}$ & $k$-points & Energy & Kinetic & Potential \\
    \hline
    \multirow{4}{*}{20}  & $\Gamma$ & -126.87466(6) & 65.493(17) & -207.492(07)\\
                         & 2x2x2    & -126.8903(08) & 65.401(11) & -207.403(09)\\ 
                         & 3x3x3    & -126.8956(10) & 65.373(11) & -207.380(09)\\
                         & 4x4x4    & -126.9058(05) & 65.325(07) & -207.356(06)\\
    \hline
    \multirow{2}{*}{40} & $\Gamma$ & -126.8729(4) & 65.436(13) & 
     -207.431(23)$\,$ \\
                        & 3x3x3    & -126.8775(4)
                        $\,$  & 65.380(12) &  -207.363(13) \\
    \hline
    \hline
  \end{tabular}
\end{table}

\subsection{Band gap}
We first discuss the band gap of LaScO$_3$, for both the experimental and DFT methods. 
From optical conductivity measurements, the spectral intensity shows a clear increase at $\approx$ 6~eV \cite{PhysRevB.48.17006}.
Numerous other optical measurements yield similar gaps.
For example, a band gap of 5.7(1)~eV for LaScO$_3$ was found from samples grown with molecular beam deposition and pulsed laser deposition \cite{doi:10.1063/1.1829781} as well as $>$~5.8~eV using epitaxial films \cite{Heeg2006}.
From this, it is clear that the band gap is nearly 6.0~eV or perhaps slightly above, with an experimental
uncertainty of a few tenths of eV. 
Previous LDA calculations underestimate the bandgap as 3.98 eV
\cite{RAVINDRAN2004554}, whereas PBE in the experimental structure has been
found to be 3.81 eV \cite{PhysRevB.86.235117}, which is clearly related to the
well-known band gap underestimation in DFT. 
This is in close agreement with our independent PBE calculations, shown in Fig. \ref{fig:pbe}, which gives a band gap of 4.04~eV. 
Note that although the settings are not identical since we employ different
pseudopotentials, gaussian basis sets vs. plane wave basis sets, slightly
different experimental geometries to the ones used previously and different codes, the results are very close.
Additionally, we provide both HF and PBE0 band structures with
25\% exact exchange, labeled as PBE0, which show the increase in the bandgap
as exact exchange weight is increased as expected, with plots in Fig. \ref{fig:bandstructure}. 
A similar behavior was demonstrated in \cite{PhysRevB.86.235117} using the HSE06 hybrid functional. 
Clearly, the band gap from the hybrid functional depends on the mixing and therefore it requires an educated guess and fine tuning in order to find proper effective exchange mixing. 

In order to probe for the accuracy and impact of the nodal surface on the band gap, we first estimate the optical
gap, with simple promotion of an electron in the Slater determinant from
VBM to CBM, which are both at the $\Gamma$ point (Fig.
\ref{fig:bandstructure}). 
In Fig. \ref{fig:qmcpack_gaps}, we show the calculated ground and excited states
from fixed-node DMC using the QMCPACK code with total of 20 atoms in the simulation cell. By varying the Fock exchange in PBE0
functional, we generate orbitals for the Slater determinant and therefore scan the corresponding nodal 
surfaces for both the ground state and excited state \cite{kolorencprb}.
Fig. \ref{fig:qmcpack_ccECP} shows the ground state and excited state calculations using the ccECPs for Sc and O and the dft-kb pseudopotential for La,
whereas the Fig. \ref{fig:qmcpack_jaron} shows the ground state and excited state calculations using the softer dft-opt pseudopotentials.
Irrespective of the pseudopotential sets used, the total energies for the ground and excited state show similar behavior. 
The minima of the DMC energy occurs using nodal surfaces of PBE0$_{30}$. 
From this, we can estimate the gap for each; the uncorrected gap using the ccECPs is 6.28(7)~eV whereas using the designed QMC potentials we estimate 6.36(7)~eV showing an excellent consistency of the direct energy differences. 
\begin{table}[!ht]
    \centering
    \caption{DMC total energies of the ground and excited state calculations using the $\Gamma$ point as the twist vector from \textsc{Qwalk}. This choice of twist accommodates both the CBM and VBM in the simulation cell, and define the excitation. The excited state calculations here use an independently optimized Jastrow for the ground and excited state and the gaps are not corrected for finite size effects. The total energies in the range from PBE0$_{20}$ to PBE0$_{30}$ range are roughly within two error bars and imply the uncorrected average gap here of about $6.15$ eV.}
    \begin{tabular}{c | ccc}
        \hline\hline
    Nodal Surface & $|\Psi_0\rangle$ & $\hat{c}^\dagger_{\rm CBM,\Gamma,\uparrow} \hat{c}_{\rm VBM,\Gamma,\uparrow} |\Psi_0\rangle$ & $\Delta_{\rm op}(\Gamma)$ \\
        \hline
        PBE     & -507.4854(32) & -507.2535(09) & 6.31(09) \\
        PBE0-10 & -507.4967(03) & -507.2580(01) & 6.493(8) \\
        PBE0-20 & -507.4898(29) & -507.2715(15) & 5.94(08) \\
        PBE0-25 & -507.4987(22) & -507.2681(28) & 6.27(10) \\
        PBE0-30 & -507.4953(16) & -507.2696(05) & 6.14(05) \\
        PBE0-40 & -507.4978(10) & -507.2702(08) & 6.19(03)\\
        PBE0-50 & -507.4915(18) & -507.2621(16) & 6.24(07)\\
        PBE0-75 & -507.4779(37) & -507.2433(06) & 6.38(10) \\
        HF      & -507.4491(37) & -507.2126(13) & 6.43(11) \\
        \hline
        \hline
    \end{tabular}
    \label{tab:relaxed_dmc}
\end{table}

We directly compare these results to another independent set of QMC calculations using the \textsc{Qwalk} package with nodal surfaces generated using the semi-local versions of our ccECPs from \textsc{Crystal}, shown in Table \ref{tab:relaxed_dmc}.
We find similar behavior in the optimal ground state, which within error uses either the PBE0$_{25}$ or PBE0$_{30}$ nodal surfaces, whereas the excited state is PBE0$_{30}$ (within statistical error). 
From the estimated gaps using consistent nodal surfaces, we see that the gap is in close agreement to the QMCPACK calculations with independent pseudopotentials. 


As an independent check on the gap, we perform quasi-particle gap calculations.
Given the high dielectric of the material, we do not expect any contributions
from excitons, so that the quasiparticle and optical gap estimates should agree
and form independent estimates of the gap from QMC. For this purpose,
we have carried out charged state calculations with the QMC designed potentials and \textsc{QMCPACK} as well as our ccECPs with \textsc{QWalk}, shown in Table \ref{tab:qp_gaps} using PBE0$_{25}$ nodal surfaces. 

\begin{table}[!t]
    \centering
    \caption{Total energies [Ha] and band gaps [eV] estimated in 20 atom supercell from independent \textsc{Qwalk} and \textsc{Qmcpack} calculations. 
    The single-reference trial wave functions with PBE0 orbitals were used and the gaps from energy  differences do not include finite size corrections.}
    \begin{tabular}{l|cc}
        \hline\hline
        Property                  & VMC-\textsc{Qwalk} & DMC-\textsc{Qwalk} \\
        \hline
        $E_N(\Gamma)$             & -506.7667(20)& -507.4987(22)  \\
        $E_{N+1}(\Gamma)$         & -506.1680(24) & -506.9369(40) \\
        $E_{N-1}(\Gamma)$         & -507.0704(25) & -507.8305(50) \\
        \hline 
        $\Delta_{\rm qp}(\Gamma)$ &   8.02(12)    &   6.26(20)   \\
             \hline\hline
             & VMC-\textsc{Qmcpack} & DMC-\textsc{Qmcpack} \\
        \hline
        $E_N(\Gamma)$ & -508.2768(20) & -508.9050(17) \\
        $E_{N+1}(\Gamma)$ & -507.6968(18) & -508.3445(18) \\
        $E_{N-1}(\Gamma)$ & -508.5888(17) & -509.2332(17) \\
        \hline 
        $\Delta_{\rm qp}(\Gamma)$ & 7.29(10)    &   6.32(09) \\
        \hline
        \hline
    \end{tabular}
    \label{tab:qp_gaps}
  \end{table}

{\bf Extrapolation analysis.} After we find close agreement between the \textsc{QWalk} and \textsc{QMCPACK}
calculations with various pseudopotentials, we study the impact of supercell sizes on the band gap and overall electronic properties. This is quite challenging due to the fact that 
the number of electrons per primitive cell is significant and therefore very large supercells would lead to prohibitively long runs. We therefore focus on calculations of the 
optical and quasiparticle gaps 
with doubled size of 40 atom cell. We collect
the results for side by side comparison in Table \ref{tab:gap-supercell} where
for the neutral 20-atom cell we use the lowest energies from Tab.\ref{tab:relaxed_dmc}. First thing to notice is that the quasiparticle and optical gaps are identical within the error bars. Furthermore, it is clear that 
 the gap estimates from plain total energy differences for each size are burdened by significant finite size biases.
 As elaborated below, these raw differences reflect finite size effects
that often have complicated dependence 
on size and shape of the supercell, differing convergence to asymptotic values for kinetic and potential energies, nature of excitations and/or net charge in the supercell.
In order 
to understand these issues and corresponding data, we write the total energies of supercells with $N$ LaScO$_3$ units for ground and optical excitation states as energy per chemical unit 
\begin{eqnarray}
E_N/N=e_{0}+A/N +B/N^{\alpha} + ... \\
E^{ex}_N/N= e_{0}^{ex}+[E_g+A^{ex}]/N + B^{ex}/N^{\alpha} + ...
\end{eqnarray}
where $e_0,e_0^{ex}$ are energies per unit, $E_g$ is the gap and $A, ... $ are parameters. 
For our case of a large gap insulator and charge neutral supercells, we expect that the exponent $\alpha >1$. 
We consider the impact of sub-leading order terms as either minor if $\alpha\ge 2$ or, possibly, as more slowly varying \cite{Shepherd} with size so that  they are effectively lumped into the $A,A_{ex}$ terms. 
This second case of more slowly varying sub-leading order contributions
would be especially relevant for studies at intermediate sizes when calculations of larger supercells might be out of reach.
Furthermore, we also consider the excitonic electron-hole effects to be marginal as argued above and as it is also clear from Table \ref{tab:gap-supercell} below. 
In what follows, we will study the optical excitation $E_g$ that makes the analysis more straightforward due to the charge neutrality, simplifying thus the finite size effects.  
Analysis of the total energies per unit is useful since 
close to the asymptotic regime
only the first two terms are important. Furthermore, the expressions show that the gap value is present in the slope of the excited state so that it can be eventually extracted from the data.
We note that we specifically distinguish between the ground and the excited states for 
the asymptotic values of energy per chemical unit.  Although they {\em should} be the same, this is not always the case at intermediate supercell sizes due to the
finite size effects as further discussed below. Clearly, this is one potential source of the finite size bias.
 The expressions also clearly identify another possible source of difficulties that is represented by the total energy offsets $A,A^{ex}$. 
 

\begin{figure*}[!ht]
  \centering
    \begin{subfigure}
        {0.48\textwidth}
        \centering
        \includegraphics[width=\textwidth]{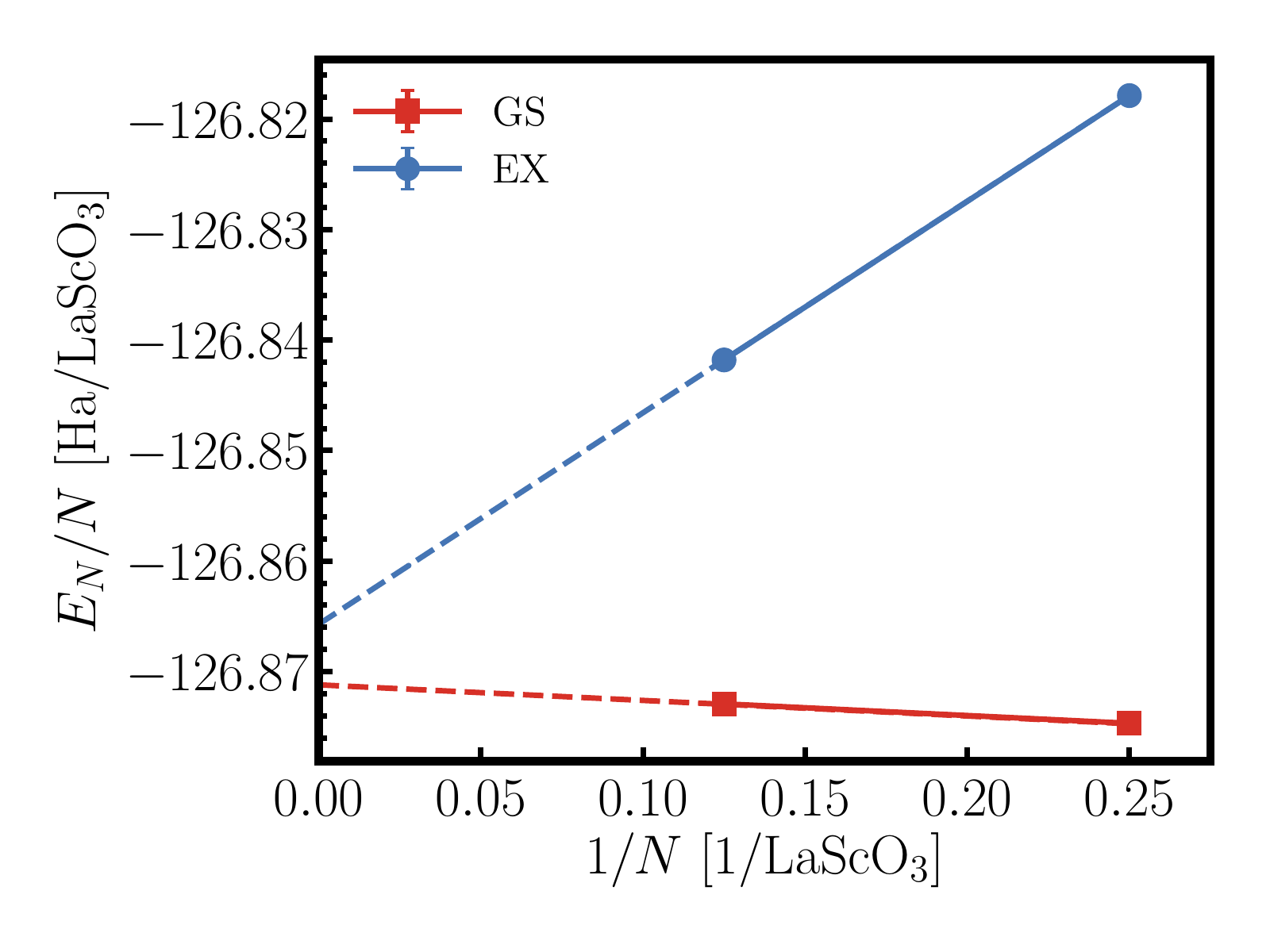}
       \caption{}
        \label{fig:extrap1}
    \end{subfigure}
    \begin{subfigure}
        {0.48\textwidth}
        \centering
        \includegraphics[width=\textwidth]{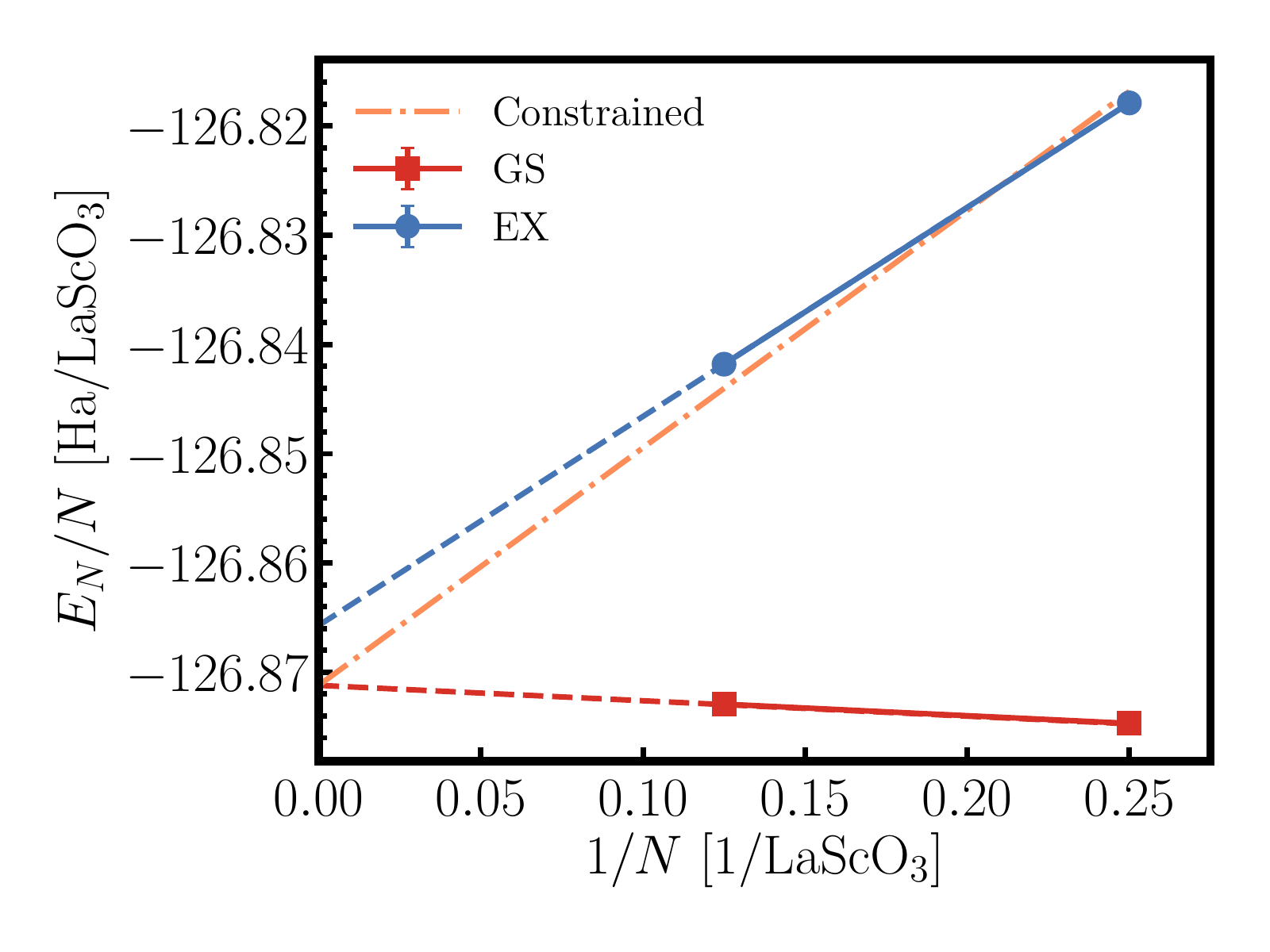}
        \caption{}
        \label{fig:extrap2}
    \end{subfigure}
    \caption{a) Extrapolations of ground (squares) and excited (circles) state energies per unit using 20- and 40-atom supercells data. Note that the extrapolated energies do not agree by a significant margin since the corresponding error bars are about 1 mHa/unit (ie, smaller than the data symbols). b) We add the constrained extrapolation of the excited state with the asymptotic
    value identical to the ground state.
    As explained in the text, any asymptotic value within a reasonable energy range
    (eg, plotted vertical axis range) that is applied consistently to both state extrapolations leads to the same gap value.}
    \label{fig:extrapolations}
\end{figure*}

Let us now outline how we address these  complications and also propose
an approach on how to extract the gap  from our data set. 

{\em i) Total energy offsets.} We first address the asymptotic
energy offsets, ie, $A$ and  $A^{ex}$, by considering 
that they do not necessarily have 
the same value. Although the offsets should nominally vanish for $N\to \infty$, it is well-known that the conditionally convergent Ewald sums can exhibit finite, ${\cal O}(1)$ terms, in the thermodynamic limit (e.g., from nonvanishing multipole moment densities, such as quadrupole moments). Similarly, the kinetic energy could converge slowly for metals \cite{foulkes2008} or in the vicinity of critical points in electronic phase diagrams, etc.
Related effects were addressed by corrections using real or reciprocal space formulations \cite{
PhysRevB.98.075122,yang2019electronic,chiesa,Shepherd}.
  Furthermore, any contribution of
sub-leading terms would effectively add to $A,A^{ex}$ within our range of sizes as mentioned above\cite{Shepherd}.
 The key observation for our data is that  the ground state value of $A$ appears to be rather small, see Table \ref{tab:gap-supercell} and 
Fig. \ref{fig:extrap1}. 
We assume that this reflects a gradual convergence due to
appreciably large number of electrons/states already in the primitive cell and rather flat bands in comparison with the expected size of the band gap. 
On the other hand, the excited state slope is an order of magnitude larger than for the ground state. Therefore we expect that $A^{ex}$ contained in the excited state slope has a similar value so that it does not overshadow the gap ``signal". Assuming that $A\approx A^{exc}$ is approximately true, we can cancel out most of this bias using the difference $A-A^{ex}$. We then expect that the remaining error would then be only a fraction of the ground state slope or perhaps even smaller. The impact of such residual error on the gap estimation would then be below our statistical uncertainty. Note that we have seen similar behavior in a previous study of phosphorene that showed essentially converged ground state thus enabling accurate gap estimation \cite{frank_many-body_2019}.

We note that this assumption might not always be valid. A possible situation where such assumption could fail can occur if the ground state slope happens to be large and comparable to the excited state slope. Such complication 
can make the gap estimation difficult or biased. Obviously, this would require further data for larger sizes 
and/or detailed analysis of the sources of biases involved.  

{\em ii) Cohesion differences.} Biases in the 
asymptotic values $e_{0}, e_{0}^{ex}$ could influence the results even more significantly. Although these values should nominally agree, in many calculations they still differ by a statistically significant margin suggesting an apparent thermodynamic limit inconsistency.
For our case, see Table \ref{tab:gap-supercell} and Fig. \ref{fig:extrap1}, the corresponding limits differ by at least six standard deviations. Regardless what is the source of the disagreement, if that would persist even for larger supercells the resulting biases would be difficult to control.
It is then inevitable that the band gap estimations from raw differences of total energies lead to a bias and that, contrary to the intuition, its influence on total energy differences can even grow with the supercell size. Although
our data indicates a seemingly ``small" difference (i.e., well below 1 mHa/unit for 40 atom system), for the total energies it is indeed very significant. 
We address this problem by an approximate {\em cohesion consistency condition} that requires both extrapolations
to converge to the same asymptotic value, 
which we label as $e_0^{asym}$. This effectively restricts
$e_0=e_0^{exc}=e_0^{asym}$
and forces both extrapolations to be consistent
in the thermodynamic limit. The corrected value of the gap is then estimated from the difference of the corresponding slopes. 
For simplicity, let us first assume that $e_0^{asym}$ is bounded by the given data range 
\begin{equation}
e_0 \leq e_0^{asym} \leq e_0^{ex}.
\end{equation}
In Fig. \ref{fig:extrap2}, we use  the  ground  state  value  of $e_0$ (that is intuitively perhaps the most appropriate choice) as the common asymptotic limit. 
Interestingly, the condition above can be relaxed and the true asymptotic value could possibly be outside this bound (note that the considered data is  
based on single $k-$point occupation). The construction is
very robust in this sense since the corresponding gap estimation does not depend on the precise value of $e_0^{asym}$. Indeed, any value from a reasonable energy range gives the same gap (basically, due to the conserved sum of angles in a triangle so that the {\em difference} between the slopes does not change). 
From this analysis we therefore conclude that the lack of consistency of energy/unit asymptotic values does play a significant role in biasing the total energy differences. Considering broader implications, if the difference between these asymptotic values is unacceptably large it simply indicates that calculations of larger supercells are still necessary in order to obtain a reliable result. 
In our experience so far, if the ground state varies mildly with the cell size, restoring the convergence of both extrapolations to the same asymptotic cohesion removes most or at least a major part of the finite size bias.  
(We have additional example(s) where essentially perfect agreement in extrapolated cohesions is achieved directly from the data and indeed the gap
estimation is then more
straightforward, to be published elsewhere.) 


\begin{figure}[!t]
  \centering
  \includegraphics[width=0.5\textwidth]{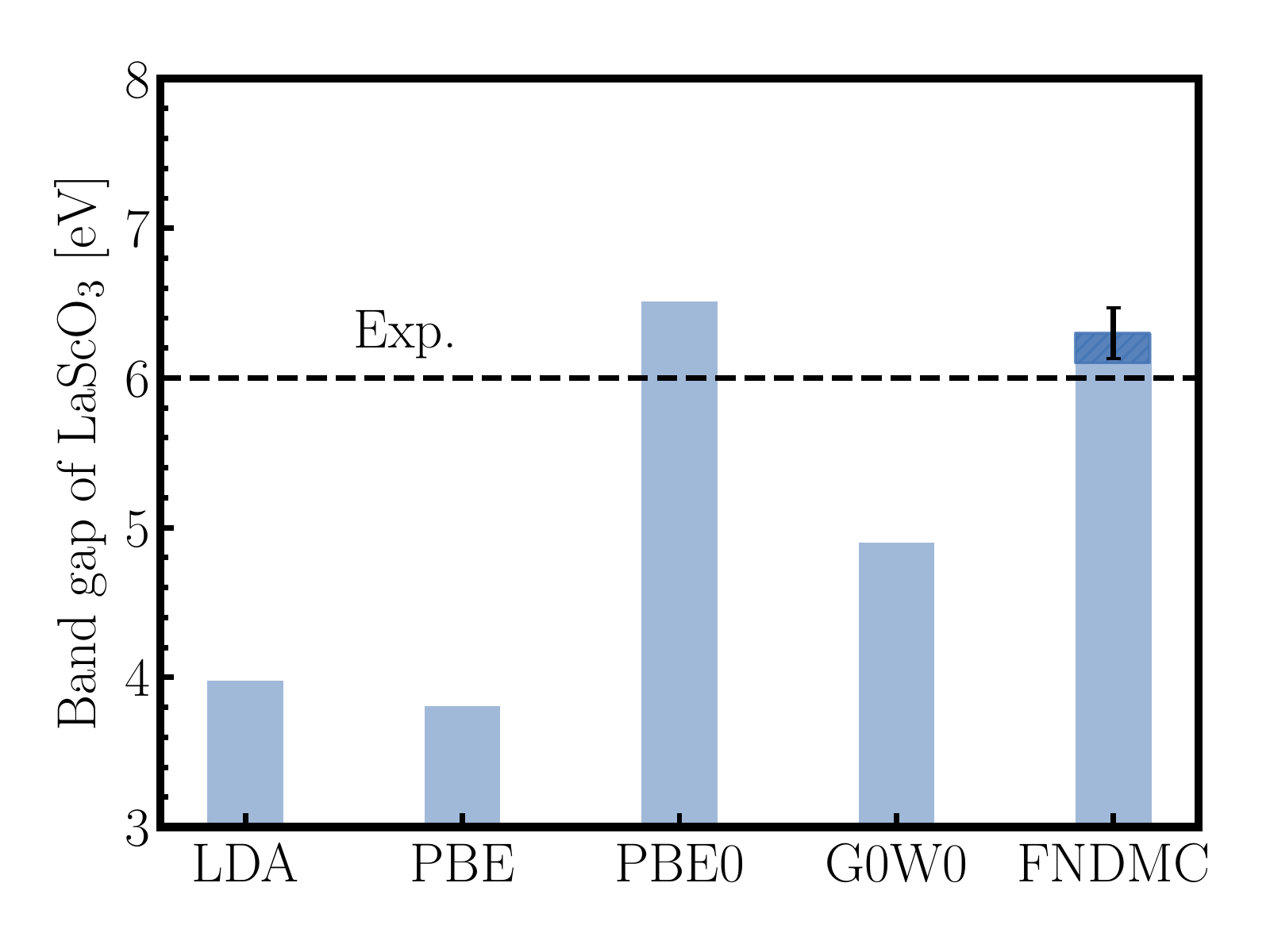}
  \caption{Band gaps of LaScO$_3$ from LDA\cite{RAVINDRAN2004554}, PBE\cite{PhysRevB.86.235117}, PBE0 (this work), $G_0W_0$ \cite{PhysRevMaterials.2.024601}, and FNDMC (this work) compared with the experimental estimation. FNDMC is given as 6.30(17)~eV, whereas the estimated of $\approx$ 0.2~eV fixed-node bias is indicated by the shaded region. 
  }
  \label{fig:gaps}
\end{figure}

Our consistency
construction leads to the gap estimation of 6.30(17) eV. In order to complete the considerations of systematic biases,  we conjecture that the fixed-node bias is between $0.1-0.2$ eV,
based on experience from previous calculations \cite{kolorencprb}. Taking this into account, the resulting best guess with all (known) systematic and random errors leads to estimate of 6.10 - 6.20 (17) eV, 
that is indeed very close to the experiments. We recall that the optical experimental data show onset of absorption at $\approx 6$ eV at ambient temperatures.
For $T=0$ though, removing the zero point and 
thermal motion effects of would increase the measured gap at least by $0.1$ eV if not more, considering corrections from electron-phonon effects in large gap solids with similar elements  \cite{PhysRevLett.107.255501,srtio3}. In fact, the resulting agreement we see is perhaps beyond expectations from a single-reference trial wave function DMC method and corroborates the previous findings that have produced accurate values for cohesive and low-lying excitations of solids. A comparison of gap estimations to experiment for various electronic structure methods is shown in Fig. \ref{fig:gaps}.

The presented construction is approximate and clearly has certain applicability limits. In particular, the data should be close to the thermodynamic limit
with the ground state close to being converged in the first place.
If this is not the case and the slopes for the ground and excited states are very similar and possibly large, the ``gap signal" might be simply lost in the statistical noise and/or present systematic biases. Biases could be caused if the offsets vary from size to size due to non-cubic cell shapes or very different aspect ratios. Although the aspect ratio changes in our case of a 20 vs. 40 atom supercell, we rely on the key feature of our band structure with a large gap and relatively flat bands. However, in general further complications could stem from slow convergence of energy components or other features such as charges, defects, etc. and may prevent reliable estimations. Further issues 
could emerge for systems with very small gaps and/or metals. Therefore, for such applications further study of finite size effects at realistic supercells sizes is highly desirable.

\begin{table}[!ht]
  \centering
  \caption{
  FNDMC total energies [Ha] calculations of the various states used to estimate the
    gaps from raw energy differences $\Delta_{\rm op},\Delta_{\rm qp}$ [eV]. We perform each calculation for the supercell
    $\Gamma$ point, which allows us to directly target the CBM and VBM in each
    supercell. There is essentially perfect agreement between optical and fundamental gaps that suggest no presence of excitonic signals. We provide the total energies and calculate the gaps from the direct energy differences,
    that are not corrected for finite size effects.}
  \label{tab:gap-supercell}
  \begin{tabular}{c|c|c}
    \hline\hline
    State/Cell & 20 atom & 40 atom \\
    \hline
     $E_N(\Gamma)$    & -507.4987(22) & -1014.9833(28)\\
    $E_{N+1}(\Gamma)$  & -506.9369(40) & -1014.4147(09) \\
    $E_{N-1}(\Gamma)$  & -507.8305(50) & -1015.3112(19) \\
    $E_N^{exc}(\Gamma$)   & 
    -507.2715(15)  & -1014.7344(22)\\
    \hline
    $\Delta_{\rm qp}$ (eV)  & 6.26(20) & 6.79(32)\\
    $\Delta_{\rm op}$ (eV)  & 6.18(08) & 6.89(23)\\
    \hline
    \hline
  \end{tabular}

\end{table}

\begin{table*}[!ht]
\renewcommand{\arraystretch}{1.5}
  \centering
  \caption{
  FNDMC energies [Ha] per LaScO$_3$ for ground and excited states used to eliminate the finite size effects from the
    estimation of the optical gap [eV] from differences ($\Delta$) as well from 
   the slopes ($E_g$ ) with finite size 
   correction.
    }
  \label{tab:gap-extrap}
  \begin{tabular}{c|c|c|c}
      \hline\hline
    State/Cell & 20 atoms & 40 atoms & $e_0$ + const$/N$ \\
    \hline
    $E_N(\Gamma)/N$ & -126.87468(55) &
    -126.87292(36) & -126.8712(12)  $-0.0141(51)/N$  \\
     \hline
     $E_N^{exc}(\Gamma)/N$ & -126.81788(38) &
    -126.84180(28) & -126.86572(74)  +0.1914(37)/$N$ \\
          \hline
    $E_N^{exc}(\Gamma)/N$ & \multirow{2}{*}{-126.81788(38)} &
    \multirow{2}{*}{-126.84180(28)} & \multirow{2}{*}{-126.8712(12) $+0.2175(37)/N$} \\   
      $e_0^{exc}\leftarrow e_0$ & & 
      &  \\
      \hline
      \hline
      Gap &  & & $E_g$ (finite size corrected)\\
      \hline
          $\Delta_{\rm op}$ (eV)  & 6.18(08) & 6.89(23) &  6.30(17) \\
    \hline\hline
  \end{tabular}
\end{table*}


{\bf Discussion.} We initiate the discussion by recalling some of the previously published results. As it is well-known DFT methods have their limitations and in particular the description of excitations can lead to rather mixed results. One of the reasons is the tendency to smooth out and distort
the excited states by  
systematic deficiencies in approximate functionals
(such as the lack of self-interaction correction) that typically leads to underestimated 
band gaps. Extensions to DFT such as hybrid functionals, through inclusion of some of the exact
exchange, correct for part of these errors and often result in rather close
agreement with experiments even with little fine-tuning. Hybrid functionals are often
reasonable using their default values of exact exchange mixing, and previous
calculations with HSE find a bandgap of 5.73~eV \cite{PhysRevB.86.235117},
whereas our independent calculations with PBE0 find a bandgap of roughly 6.51~eV
using the default parameters. While these estimates are not perfect, they are
relatively close to the experimental values. Improvements for hybrids can be
obtained by retuning the exact exchange parameter, however this requires one to
have experimental data to compare against and is not purely {\it ab initio}.
Many-body perturbation theory such as $G_0W_0$ is often used to improve the DFT
band gap issues. However, the results can be rather sensitive to the
underlying DFT orbitals \cite{zhao19}. A very systematic study using $G_0W_0$
was carried out for a wide variety of ABO$_3$ compounds, and for the majority
the calculated band gaps largely agree with experiment. However
for LaScO$_3$ the $G_0W_0$ bandgap is estimated
to be between 4.5 and 4.9~eV \cite{PhysRevMaterials.2.024601}. This
discrepancy  with experiments is justified by stating: ``LaScO$_3$ is a clear band insulator with the
first (charge-transfer) optical excitation arising from the O$-p$ to Sc$-d$
transition; the experimental spectrum does not clearly show the tail at the
bottom of the spectrum well visible in $GW$. We believe that the $GW$ band
structure is reliable in this respect, and that the onset of optical absorption
is not easily detected in the experiment. We therefore trust that our predicted
band gap of 4.9~eV should be more reliable than the experimental estimate of
6.0~eV.'' We would like make the following comments on two aspects of this interesting material. One point to consider is that the $3d$-channel is not the only candidate for the lowest excitations stemming mainly from the Sc atom. The actual situation might be more complex and probably involves a number of channels and many-body effects. In this respect, there is a rather unique example of bounded atomic Sc$^{-}$ anion for which one expects the ground state to be a triplet with additional electron occupying the $d$-orbital, ie, 
$^{3}F(3d^24s^2)$. Surprisingly, this is incorrect and such a state is not bounded \cite{mitas1992,hotop}.
The true Sc$^-$ bounded lowest state actually does not conform to the Hund's rules; it is a singlet and the additional electron goes to the $4p$ state, i.e., 
it is $^1D(3d4s^24p)$, showing nontrivial 
correlations and competition of several one-particle atomic channels. We believe that 
such correlations might be quite difficult to describe even by sophisticated and precise perturbational approaches
and indeed many-body wave functions 
are crucial for describing such effects \cite{mitas1992}.

Furthermore, we also consider the fact that absorption data could be challenging to analyze when the band gap is
large. However, the underestimation from $GW$ using PBE orbitals does not necessarily imply the stated claim,
given the significant variability of typical $GW$ results with respect to their
underlying orbital sets \cite{zhao19}. 
Our DMC results indicate that the gap, at least in the ideal system without defects or non-stoichiometries, appears to be around or marginally above $\approx$ 6 eV, in an excellent agreement
with several published spectra. Therefore we are inclined to accept the optical experiments
as representative in this regard. High quality direct and inverse photoemission experiments for LaScO$_3$
would be highly desirable
in corroborating the results of both our calculations as well as the data from existing optical measurements.

\section{Conclusions}
We present a quantum Monte Carlo study of LaScO$_3$ based on real space sampling and fixed-node diffusion Monte Carlo method.  Experiments show that this scandate is a large gap nonmagnetic insulator as
has been consistently reconfirmed by several experiments.
Despite the fact that optical data is
often more complex to analyze, for large
dielectric materials the lowest excitations should effectively measure the fundamental gap due to absence of strong excitons. In this Mott-Wannier limit, the
interaction between electrons and holes is largely screened, and the resulting
excitons are delocalized with a very small binding energy. 

Our calculations are based on standard the DMC workflow and we are able to obtain accurate band gaps when compared to experiment, without the need for fine-tuning any non-variational parameters.
Other electronic structure methods rely on guidance from experiments, such as adjusting the exact exchange mixing parameter in hybrid functionals or inserting Hubbard $U$ in DFT+$U$ approaches. Similarly, changing the initial orbitals sets, i.e., the starting point bias
in $GW$ methods can affect and possibly bias the predictions as well.  As an alternative, diffusion Monte Carlo relies only on the variational principle and on the quality 
of the nodal set 
of the explicitly constructed many-body trial wave function.
Note that the usual issue of finite basis set in correlated wave function methods does not arise in DMC since the random walks represent sampling of the complete position space.
We show that for systems such ours, 
the standard ansatz for excited states such as promotion from the valence to the conduction band works quite well when coupled with the nodal surface optimization using one-parameter hybrid DFT effective 
Hamiltonian.

We have carried out also a rather detailed study of finite size effects such as twist averages for the proper estimation of cohesion. In fact, our cohesive energy is a genuine prediction since to the best of our knowledge that quantity has not been measured yet. Similarly, we have studied  the finite size biases in estimations of the band gap for different supercell sizes. Since our system exhibits
relatively flat bands when compared with the gap, we have encountered rather mild finite size dependence of the ground state at intermediate sizes. We then showed that it is important to take into consideration both the ground and excited states scaling to large sizes
so as to guarantee the consistency in thermodynamic limit. Using this as a consistency condition has enabled us to conclude that the band gap is located at 6 eV and possibly marginally higher. Based of previous transition
metal oxide QMC calculations we conjecture that the bias 
from the fixed-node approximation is 
around 0.1 to 0.2 eV
which is essentially on par with our statistical resolution.

Although our agreement with the experiment is very encouraging we would like to reflect on the origin of systematic errors still involved.    We recapitulate the following possible sources:

a) Finite size effects, which we treat approximately, are still very important for further considerations. This concerns both the offsets and sub-leading order terms that could be still significant at intermediate sizes. Within the presented data the gap estimation is essentially locked by the introduced consistency condition
at the asymptotic limit, however, for other quantities more studies and insights are necessary.

b) Residual fixed-node errors that remain after taking differences are equally relevant. The goal is to get the resulting fixed-node bias to such level that it would not affect the differences of interest within the desired statistical accuracy. Obviously, for larger systems the cost and difficulty of such exploration can be very very challenging.

c) Neglect of zero point and thermal motion effects on the band gap and possibly other quantities of interest. 
For these effects we roughly estimate that their impact on the band gap is of the order of 0.1 eV. Here we consider some of the recent results \cite{srtio3,gonze} for transition and simple metal oxide materials such as SrTiO$_3$,
TiO$_2$ and SnO$_2$ where phononic corrections are in the range 
 $\approx$ 0.1 and 0.3 eV. This clearly requires further study.

In order to cross-validate a number of technical aspects of the QMC methodology,  most of the calculations have been carried out along two independent tracks. One used the 
\textsc{Qwalk} code with gaussian basis sets, correlation consistent ECPs (ccECPs), and the \textsc{Crystal} code for generating the self-consistent gaussian-based solid state orbitals. The other track employed 
DFT-generated pseudopotentials, \textsc{Quantum Espresso} code with plane wave basis
and \textsc{Qmcpack} for QMC runs.
Overall, we find excellent agreement between the results using these technically very different paths, independent codes,
effective Hamiltonians and basis sets.
These two sets of results provide  a systematic corroboration of the robustness of the QMC methods and tools for calculations of transition metal oxide systems and other complex materials in future.

For more complicated systems and further gains in accuracy, systematic improvement of the ground and excited states may be necessary \cite{zhao19}.
In addition, better understanding of finite size effects for realistic sizes of supercells will be also an important subject of future studies. Nevertheless, the methods presented here provide an excellent starting point and a first step towards more accurate and further elaborated {\it ab inito} calculations where more sophisticated trial wave function ansatz may be necessary, beyond just the Slater-Jastrow form.

\bigskip

{\bf Acknowledgements} 

This work presents two sets of quantum Monte Carlo results, methodologies, packages and calculations. The work based on the QMCPACK package and associated tools ($\approx 50$\% of the project) has been supported by the U.S. Department of Energy, Office of Science, Basic Energy Sciences, Materials Sciences and Engineering Division, as part of the Computational Materials Sciences Program and Center for Predictive Simulation of Functional Materials.

The authors would like to thank Luke Shulenburger for reading the manuscript and for comments.

The presented work that used the QWalk package and corresponding tools ($\approx 50$\% of the project)
has been supported by the U.S. Department of Energy,
Office of Science, Basic Energy Sciences (BES) under the award de-sc0012314.

The calculations were performed at Sandia National Laboratories and at TACC under XSEDE.

Sandia National Laboratories is a multi-mission laboratory managed and operated by National Technology and Engineering Solutions of Sandia LLC, a wholly owned subsidiary of Honeywell International, Inc. for the U.S. Department of Energy's  National Nuclear Security Administration under
Contract No. DE-NA0003525.

This paper describes objective technical results and analysis. Any subjective views or opinions that might be expressed in the paper do not necessarily represent the views of the U.S. Department of Energy or the United States Government.

\bigskip 
\bigskip

\bibliography{main.bib}

\end{document}